\documentclass[12pt,a4paper]{article}
\usepackage[T2A]{fontenc}
\usepackage{epsfig}

\begin{document}

\title{A HIGH STATISTICS STUDY OF THE BETA-FUNCTION IN THE SU(2) LATTICE THERMODYNAMICS}
\author{ S.~S.~Antropov\thanks{E-mail:santrop\_2@yahoo.com}, V.~V.~Skalozub\thanks{E-mail:skalozubv@daad-alumni.de}\\
{\small Oles Honchar Dnipropetrovsk National University,}\\
{\small Dnipropetrovsk, Ukraine}\\
O.~A.~Mogilevsky\thanks{E-mail:mogilevsky.oleg@gmail.com}\\
{\small Bogolyubov Institute for Theoretical Physics of the National Academy }\\
{\small of Sciences of Ukraine, Kiev, Ukraine}}
\date{}
\maketitle
\begin{abstract}
 The beta-function is investigated on the lattice in $SU(2)$ gluodynamics. It is determined within a scaling hypothesis while a lattice size fixed to be taken into account. The functions calculated are compared with the  ones obtained in the continuum limit. Graphics processing units (GPU) are used as a computing platform that allows gathering a huge amount of statistical data. Numerous beta-functions are analyzed for various lattices. The coincidence of the lattice beta-function and the analytical expression in the region of the phase transition is shown. New method for estimating a critical coupling value is proposed.
\end{abstract}

\section{Introduction}

\hskip 0.5 cm The beta-function is one of the main objects in quantum field theory. It defines scaling properties of the theory in different regions of dynamic variables. It is defined as
\begin{eqnarray}\label{antrop_2}
\beta_f(g_\mu)=\mu^2\frac{\partial \overline{g}(\mu^2)}{\partial (\mu^2)},
\end{eqnarray}
where $\beta_f(g_\mu)$ is the beta-function, $g_\mu\equiv\overline{g}(\mu^2)$ -- the effective coupling constant, $\mu$ -- the normalizing momentum.

For the case of the Monte-Carlo (MC) calculations in $SU(N)$ lattice gluodynamics the beta-function has the form
\begin{eqnarray}
\beta_f(g)=-a\frac{d g}{d a},
\end{eqnarray}
where $a$ replaces the parameter $\mu^2$, $a$ - is the lattice spacing. Lattice spacing is a free parameter of the theory. In particular, the calculation of $\beta_f(g)$ is one of the ways to define $a$.

In analytical approach, the beta-function is well described by an expansion as power series of coupling constant. In the cases of quantum chromodynamics or $SU(N)$ lattice gluodynamics, a non-perturbative beta-function attracts the most interest.

In ref. \cite{antrop_Mogilevsky:2006} a new special method was developed. Namely, the effects connected with the final sizes of a lattice were taken into account, and scaling near the critical point of $SU(N)$ lattice gauge theories has been  considered without attempt to reach a continuum  limit.


\hskip 0.5 cm The goal of the present paper is the detailed investigation  and development of this approach. In $SU(2)$ gluodinamics, we calculate the beta-functions on different lattices and  compare  their values with those obtained in a continuum limit.

\section{Analytical expression}

\hskip 0.5 cm The beta-function describes the dependence of the lattice spacing $a$ on a coupling constant $g$
\begin{eqnarray}\label{antrop_1}
\beta_f(g)=-a\frac{d g}{d a}.
\end{eqnarray}
Our calculations are based on the special form of the definition of the  beta-function \cite{antrop_Mogilevsky:2006}. Let us consider a transformation
\begin{eqnarray}
a\rightarrow a'=ba=(1+\Delta b)a.
\end{eqnarray}
Under this transformation the definition (\ref{antrop_1}) becomes
\begin{eqnarray}\label{antrop_opredelen}
-a\frac{d g}{d a}=-\lim_{b \to 1}\left(a\frac{g(ba)-g(a)}{ba-a}\right)=
-\lim_{b \to 1}\frac{d g}{d b}=\beta_f(g).
\end{eqnarray}

The singular part of the free energy density can be described by the universal finite-size scaling function \cite{antrop_Fingberg:1992ju}
\begin{eqnarray}
f(t,h,N_\sigma,N_\tau)=\left(\frac{N_\sigma}{N_\tau}\right)^{-3}Q_f\left(g_t\left(\frac{N_\sigma}{N_\tau}\right)^{1/\nu},
g_h\left(\frac{N_\sigma}{N_\tau}\right)^\frac{\beta+\gamma}{\nu}\right),
\end{eqnarray}
where $\beta, \gamma, \nu$ are the critical indexes of the theory.   Due to the finite size scaling hypothesis, these indexes coincide with the critical indexes of 3-d Ising model. The scaling function $Q_f$ depends on the reduced temperature $t=\frac{T-T_c}{T_c}$ and on the external field strength $h$ through the thermal and magnetic scaling fields
\begin{eqnarray}\label{antrop_coeff}
g_t &=& c_t t(1+b_t t),\\\nonumber
g_h &=& c_h h(1+b_h t)
\end{eqnarray}
with non-universal coefficients $c_t, c_h, b_t, b_h$ which are still carrying a possible $N_\tau$ dependence.

The existence of the scaling function $Q$ \cite{antrop_M. N. Barber,antrop_V. Privman} allows developing a procedure to renormalize the coupling constant $g^{-2}$ by using two different lattice sizes $N_\sigma, N_\tau$ and $N'_\sigma, N'_\tau$ ($N_\sigma$ is the number of lattice nods in spatial directions, $N_\tau$ -- the number of lattice nods in time direction). Let us fix $\frac{N'_\tau}{N_\tau}=\frac{N'_\sigma}{N_\sigma}=b$ and perform a scale transformation
\begin{eqnarray}
a&\rightarrow& a'=ba,\\\nonumber
N_\sigma&\rightarrow& N'_\sigma=\frac{N_\sigma}b,\\\nonumber
N_\tau&\rightarrow& N'_\tau=\frac{N_\tau}b.
\end{eqnarray}

Then the phenomenological renormalization is defined by the following equation
\begin{eqnarray}\label{antrop_transformation}
Q(g^{-2},N_\sigma,N_\tau)=Q\left((g')^{-2},\frac{N_\sigma}b,\frac{N_\tau}b\right).
\end{eqnarray}
It means that the scaling function $Q$ remains  unchanged if the lattice size is rescaled by a factor $b$ and the inverse coupling $g^{-2}$ is shifted to $(g')^{-2}$ simultaneously. Taking the derivative with respect to the scale parameter $b$ of the both sides of (\ref{antrop_transformation}) and using (\ref{antrop_opredelen}) we obtain the expression
\begin{eqnarray}\label{antrop_betaf}
a\frac{d g^{-2}}{d a}=\frac{\frac{\partial Q(g^{-2},N_\sigma,N_\tau)}{\partial ln N_\sigma}+\frac{\partial Q(g^{-2},N_\sigma,N_\tau)}{\partial ln N_\tau}}{\frac{\partial Q(g^{-2},N_\sigma,N_\tau)}{\partial g^{-2}}}.
\end{eqnarray}

Fourth derivative of $f$ in $h$ taken at $h=0$ and divided by $\chi^2(\frac{N_\sigma}{N_\tau})^3$ is called the Binder cumulant \cite{antrop_Binder:1981zz}
\begin{eqnarray}
g_4=\frac{\frac{\partial^4 f}{\partial h^4}}{\chi^2(\frac{N_\sigma}{N_\tau})^3}\Biggm|_{h=0}.
\end{eqnarray}
It identically coincides with the scale function \cite{antrop_Binder:1981zz}
\begin{eqnarray}\label{antrop_cummul}
g_4=Q_{g_4}\left(g_t\left(\frac{N_\sigma}{N_\tau}\right)^\frac1\nu\right).
\end{eqnarray}
Binder cumulant $g_4$ is calculated through the Polyakov loops on a lattice \cite{antrop_Binder:1981zz}
\begin{eqnarray}
g_4=\frac{\langle P^4\rangle}{\langle P^2\rangle^2}-3.
\end{eqnarray}

We get the expression for the beta-function
\begin{eqnarray}\label{antrop_betafu}
a\frac{d g^{-2}}{d a}=\frac{\frac{\partial g_4}{\partial ln N_\sigma}+\frac{\partial g_4}{\partial ln N_\tau}}{\frac{\partial g_4}{\partial g^{-2}}}=\frac14\frac{\frac{\partial g_4}{\partial ln N_\sigma}+\frac{\partial g_4}{\partial ln N_\tau}}{\frac{\partial g_4}{\partial\beta}}.
\end{eqnarray}

\section{Lattice observables}

\hskip 0.5 cm Let us calculate the beta-function using (\ref{antrop_betafu}). As the lattice size is discrete, it is necessary to replace the derivatives in (\ref{antrop_betafu}) by the finite differences which are calculated on lattices with the closest $N_\sigma, N_\tau$ (and corresponding $g_4(N_\sigma,N_\tau)$):
\begin{eqnarray}\label{antrop_podstav}
\frac{\partial g_4(\beta,N_\sigma,N_\tau)}{\partial ln N_\sigma}\rightarrow\frac{g_4(\beta,N'_\sigma,N_\tau)-g_4(\beta,N_\sigma,N_\tau)}{ln (\beta,N'_\sigma/N_\sigma)},\\\nonumber
\frac{\partial g_4(\beta,N_\sigma,N_\tau)}{\partial ln N_\tau}\rightarrow\frac{g_4(\beta,N_\sigma,N'_\tau)-g_4(\beta,N_\sigma,N_\tau)}{ln (\beta,N'_\tau/N_\tau)}.
\end{eqnarray}

Such replacement,
\begin{eqnarray}\label{antrop_zamena}
\frac{\partial g_4}{\partial\beta}\rightarrow\frac{\Delta g_4}{\Delta\beta},
\end{eqnarray}
leads to huge computing errors. Near the  phase transition area, the dispersion is increased  and the substitution  (\ref{antrop_zamena}) becomes  not reasonable. For different lattices investigated, the amount of  data near the critical region varies from $120$ up to $600$ points, but the error for (\ref{antrop_zamena}) still remains large.

\begin{flushright}
\textbf{Table 1.} Tested fitting curves
\end{flushright}
\begin{center}
\begin{tabular}[c]{|l|c|}\hline
 \rule{0pt}{3ex}\scriptsize
 $Function$ & $Parameters$\\\hline
 \rule{0pt}{3ex}\scriptsize
 $A_1 + \frac{A_2-A_1}{1 + 10^{(\beta_0-\beta)*p}}$ & $A_1,A_2,\beta_0,p$\\\hline
 \rule{0pt}{3ex}
 $\frac{A_1-A_2}{1 + (\frac{\beta}{\beta_0})^p}+A_2$ & $A_1,A_2,\beta_0,p$\\\hline
  \rule{0pt}{3ex}
 $\frac{A_1-A_2}{1 + e^{(\beta-\beta_0)/p}}+A_2$ & $A_1,A_2,\beta_0,p$\\\hline
\end{tabular}
\end{center}

Our the best fits (see Fig. 1, Tab. 2) are reached for the function
\begin{eqnarray}\label{antrop_fitfunc}
g_4 = A1 + (A2-A1)/(1 + 10^{(\beta_0-\beta)*p}),
\end{eqnarray}
where $A1, A2, \beta_0, p$ are the fitting parameters.

\begin{center}
\includegraphics[height=6cm,keepaspectratio]{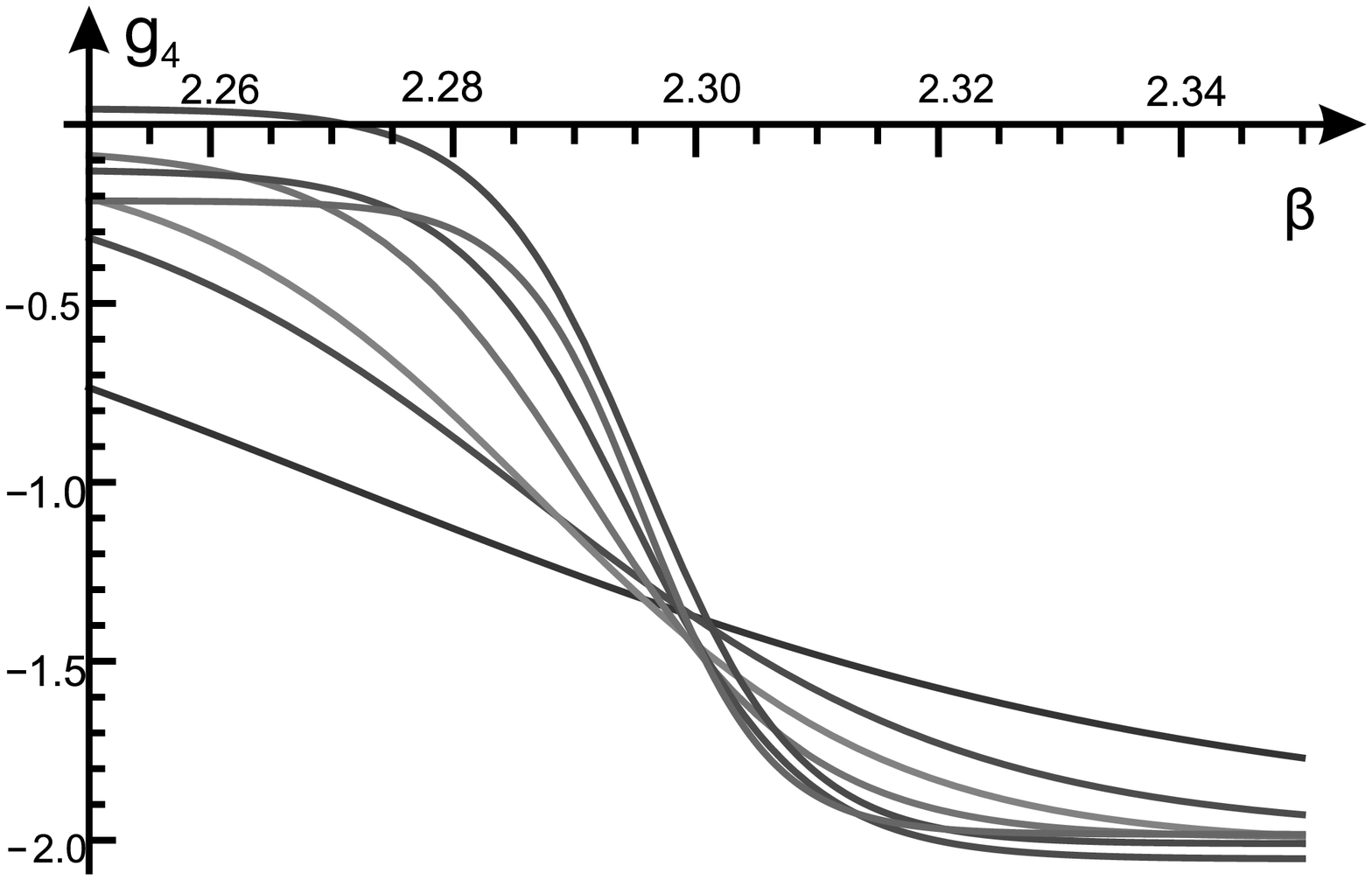}
\mbox{\parbox[t]{0.7\hsize}{{\textbf{Figure 1.} Binder cumulants. Cumulants are received on lattices with $N_\tau=4$, and $N_\sigma=8$, $12$, $16$, $24$, $28$, $32$. The higher number of nods in the lattice corresponds with the sharper step. All curves intersect each other in a local area and as it comes from the theory these curves should intersect in one point (the critical point).}}}
\end{center}

If one knows $g_4$ in an analytical form, it is possible to calculate $\frac{\partial g_4}{\partial\beta}$ straightforwardly. However, the result of $g_4$ calculations  is a set of points. To reveal a functional dependence on this sequence, it is necessary to apply some fitting procedure. For this procedure we chose the step functions, since  the critical area of $g_4$ is a steplike (see Tab. 1).

In Tab. 2 the best fits for number of lattices are represented. We have analyzed up to 600 points for some lattices and have reached small values (down to $10^{-3}$) of $\chi^2$ function.

\begin{flushright}
\textbf{Table 2.} Fitting of Binder cumulants by $A_1 + \frac{A_2-A_1}{1 + 10^{(\beta_0-\beta)*p}}$
\end{flushright}
{\scriptsize
\begin{center}
\begin{tabular}[c]{|l|c|c|c|c|c|c|c|c|}
 \cline{2-9} \multicolumn{1}{c}{~} & \multicolumn{5}{|c|}{\bf \rule{0pt}{3ex}\scriptsize \parbox{2.0cm}{\begin{center}Parameters\end{center}} } & \multicolumn{1}{c}{~} & \multicolumn{2}{|c|}{\bf \parbox{1.5cm}{\begin{center}Fitting range\end{center}}}\\
 \hline\rule{0pt}{3ex} {\bf \scriptsize \parbox{1.5cm}{\begin{center}Lattice\end{center}}} & $\chi^2$ & $A_1$ & $A_2$ & $\beta_0$ & $p$ & {\bf \parbox{1.1cm}{\begin{center}Number of points\end{center}}} & $\beta_{min}$ & $\beta_{max}$\\\hline\hline
 \rule{0pt}{3ex}
 $N_\tau=4, N_\sigma=8$  & $0.009$ & $-1.953$ & $-0.0523$ & $2.2705$ & $-12$ & $126$ & $1.7$ & $2.95$\\
 \rule{0pt}{3ex}
 $N_\tau=4, N_\sigma=8$  & $0.012$ & $-1.957$ & $-0.0507$ & $2.2747$ & $-11$ & $26$ & $1.7$ & $2.95$\\
 \rule{0pt}{3ex}
 $N_\tau=4, N_\sigma=12$ & $0.025$ & $-1.98$ & $-0.1$ & $2,286$ & $-24$ & $253$ & $1.7$ & $2.95$\\
 \rule{0pt}{3ex}
 $N_\tau=4, N_\sigma=12$ & $0.011$ & $-2$ & $-0.04$ & $2,289$ & $-16$ & $26$ & $1.7$ & $2.95$\\
 \rule{0pt}{3ex}
 $N_\tau=4, N_\sigma=16$ & $0.029$ & $-2.01$ & $-0.066$ & $2.287$ & $-30.1$ & $236$ & $1.7$ & $2.95$\\
 \rule{0pt}{3ex}
 $N_\tau=4, N_\sigma=16$ & $0.013$ & $-1.99$ & $-0.05$ & $2.292$ & $-30.9$ & $26$ & $1.7$ & $2.95$\\
 \rule{0pt}{3ex}
 $N_\tau=4, N_\sigma=20$ & $0.055$ & $-2$ & $-0.065$ & $2.291$ & $-48$ & $246$ & $1.7$ & $2.95$\\
 \rule{0pt}{3ex}
 $N_\tau=4, N_\sigma=24$ & $0.1$ & $-2.0098$ & $0.044$ & $2.296$ & $-68$ & $126$ & $1.7$ & $2.95$\\
 \rule{0pt}{3ex}
 $N_\tau=4, N_\sigma=24$ & $0.006$ & $-2.001$ & $0.061$ & $2.291$ & $-27$ & $26$ & $1.7$ & $2.95$\\\hline
 \rule{0pt}{3ex}
 $N_\tau=4, N_\sigma=28$ & $0.089$ & $-2.05$ & $-0.13$ & $2.29$ & $-62$ & $626$ & $1.7$ & $2.95$\\\hline
 \rule{0pt}{3ex}
 $N_\tau=4, N_\sigma=28$ & $0.012$ & $-1.99$ & $-8\cdot10^{-5}$ & $2.28$ & $-21$ & $26$ & $1.7$ & $2.95$\\\hline
 \rule{0pt}{3ex}
 $N_\tau=4, N_\sigma=32$ & $0.12$ & $-1.984$ & $-0.2$ & $2.3$ & $-84$ & $626$ & $1.7$ & $2.95$\\
 \rule{0pt}{3ex}
 $N_\tau=4, N_\sigma=32$ & $0.01$ & $-1.988$ & $0.014$ & $2.27$ & $-28$ & $26$ & $1.7$ & $2.95$\\
 \rule{0pt}{3ex}
 $N_\tau=4, N_\sigma=36$ & $0.19$ & $-2$ & $-0.27$ & $2.3$ & $-105$ & $600$ & $2.28$ & $2.31$\\\hline
 \rule{0pt}{3ex}
 $N_\tau=16, N_\sigma=20$& $0.094$ &$-1.17$& $-0.017$& $2.68$ & $-7$ & $126$  & $1.7$ & $2.95$\\
 \rule{0pt}{3ex}
 $N_\tau=16, N_\sigma=24$& $0.054$ & $-1.7$ & $0.04$ & $2.75$ & $-6$ & $26$   & $1.7$ & $2.95$\\
 \rule{0pt}{3ex}
 $N_\tau=16, N_\sigma=28$& $0.021$ & $-1.6$ & $-0.017$& $2.67$& $-17$ &$26$   & $1.7$ & $2.95$\\
 \rule{0pt}{3ex}
 $N_\tau=16, N_\sigma=32$ & $0.021$ & $-1.7$ & $0.03$ & $2.69$ & $-23$& $126$ & $1.7$ & $2.95$\\\hline
\end{tabular}
\end{center}
}

Now we turn to an interesting feature of these fits. Parameters of the curve, which based on 600 data points, are nearly the same as parameters (especially $\beta_0$) of the curve, which based on 25 data points. The parameter $\beta_0$ coincides (to within 2 up to 3 digits) with an inverse critical coupling constant for a corresponding lattice (see Tab. 3, ref. \cite{antrop_Fingberg:1992ju}, \cite{antrop_Velytsky:2007gj}).

\begin{flushright}
\textbf{Table 3.} Values of the inverse coupling constant
\end{flushright}
\begin{center}
{\scriptsize \small
\begin{tabular}[c]{|l|c|c|c|c|}\hline
 \rule{0pt}{3ex}\scriptsize
 $N_\tau$  & $2$     & $4$     & $6$     & $8$\\\hline
 \rule{0pt}{3ex}
 $\beta_c$ & $1.875$ & $2.301$ & $2.422$ & $2.508$\\\hline
\end{tabular}}
\end{center}

It is common to use the linear fits for critical point findings. Because of the dispersion in critical region these fits need a lot of data to be performed. Using both listed above properties one can estimate the inverse critical coupling using just few points. For more precise calculations one can use the function (\ref{antrop_fitfunc}) with data, which are from above and below critical region. The dispersion for these data is much less than for data, which are near critical area, so one need much less statistics than usually.

The expression for the beta-function in lattice variables reads:
\begin{eqnarray}\label{antrop_betafuclat}
\beta_f(\beta)=\frac1{\beta^{3/2}}\cdot\frac{\frac{g_4(\beta,N'_\sigma,N_\tau)-g_4(\beta,N_\sigma,N_\tau)}{ln (N'_\sigma/N_\sigma)}+\frac{g_4(\beta,N_\sigma,N'_\tau)-g_4(\beta,N_\sigma,N_\tau)}{ln (N'_\tau/N_\tau)}}{\frac{\partial g_4(\beta,N_\sigma,N_\tau)}{\partial\beta}}.
\end{eqnarray}
It will be used below.

\section{Calculation of the beta-function}

\hskip 0.5 cm We chose the heat-bath as working algorithm in MC procedure. We use standard form of Wilson action of the $SU(2)$ lattice gauge theory. In the MC simulations, we use the hypercubic lattice $L_t\times
L_s^3$ with hypertorus geometry.

We use the General Purpose computation on Graphics Processing Units (GPGPU) technology allowing studying large lattices on personal computers. Performance analysis ~indicates ~that the GPU-based MC simulation program shows better speed-up factors for big lattices in comparing with the CPU-based one. For the majority  lattice geometries the GPU vs. CPU (single-thread CPU execution) speed-up factor is above 50 and for some lattice sizes could  overcome the factor 100.

The plots of dependencies of the beta-function on the inverse coupling constant are shown below.

\begin{figure}[h]
\begin{minipage}[h]{0.49\linewidth}
\center{\includegraphics[width=0.9\linewidth,keepaspectratio]{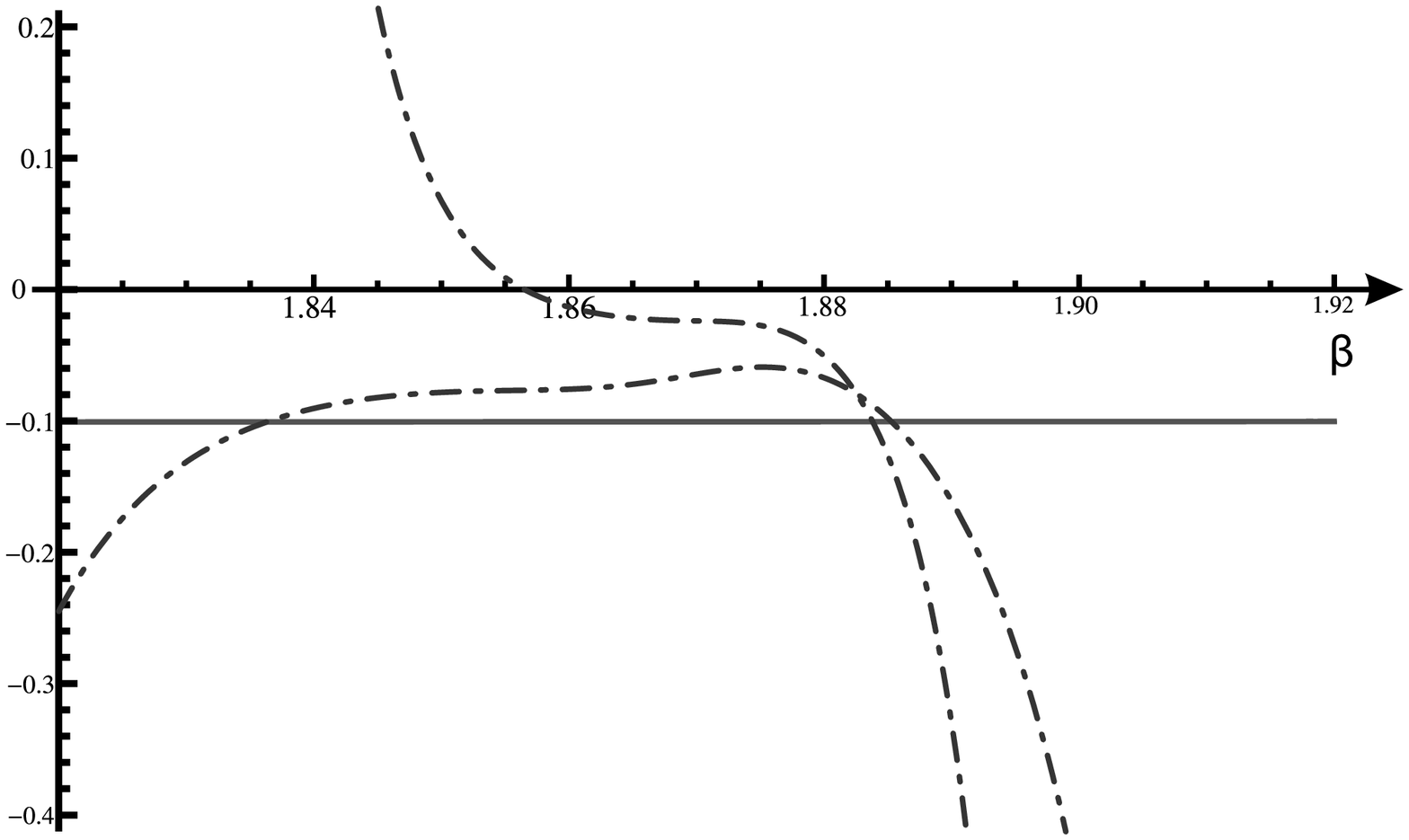}
\\ Figure 2. The solid line represents the beta-function in asymptotic expansion. Dashed lines with a point - the beta-functions (\ref{antrop_betafuclat}), $N_\tau=2$, $N_\sigma=8, 16, 20$, $\Delta N_\tau=N'_\tau-N_\tau=2$, $\Delta N_\sigma=N'_\sigma-N_\sigma=4$.}
\end{minipage}
\hfill
\begin{minipage}[h]{0.49\linewidth}
\center{\includegraphics[width=0.9\linewidth,keepaspectratio]{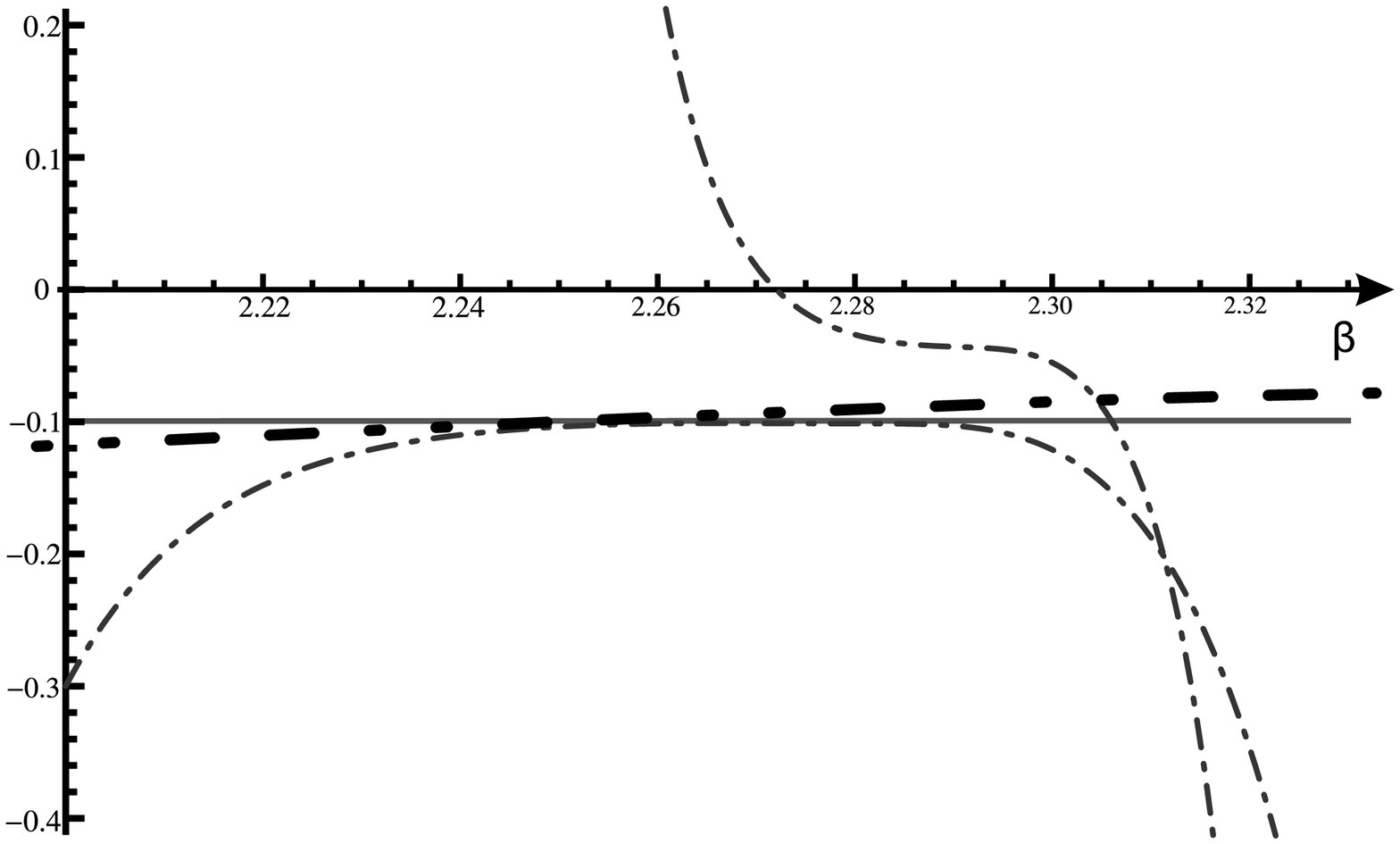}
\\ Figure 3. Same as above. Dashed lines with a point - the beta-functions  (\ref{antrop_betafuclat}), $N_\tau=4$, $N_\sigma=12, 20$, $\Delta N_\tau=N'_\tau-N_\tau=2$, $\Delta N_\sigma=N'_\sigma-N_\sigma=4$. The Dashed line with two points is the beta-function received in ref. \cite{antrop_Engels:1994xj}.}
\end{minipage}
\end{figure}

The standard deviation of the function (\ref{antrop_betafuclat}) is the smallest one near the critical point. It comes from analysis of Binder cumulants. Cumulants decrease linearly in the critical area and change little above and belove that area. Therefore $\frac{\partial g_4(\beta,N_\sigma,N_\tau)}{\partial\beta}$ in the bottom of (\ref{antrop_betafuclat}) comes to $0$ and leads (\ref{antrop_betafuclat}) to infinity. Beta-function values which are calculated near critical point are in good agreement with known results \cite{antrop_Engels:1994xj}.

\section{Conclusions}

\hskip 0.5 cm We have performed high-statistics  calculations of the beta-function in $SU(2)$ lattice gluodynamics. These calculations became possible due  to technology of GPU calculations.

The key point for our investigations is definition (\ref{antrop_opredelen}) \cite{antrop_Mogilevsky:2006}. It gives a possibility to analyze a finite size of the lattice.

We have constructed and analyzed the lattice beta-functions for a wide range of different lattices.

Values of all beta-functions in critical region are the same for different functions. In particular, the values of the beta-functions (\ref{antrop_betafuclat}) in critical region are almost the same as the values obtained in ref. \cite {antrop_Engels:1994xj}.
The fast method of determination of the inverse critical constant on a lattice based on the formula (\ref{antrop_fitfunc}) is proposed.

\end{document}